\documentclass[review]{elsarticle}

\usepackage{hyperref}

\journal{Neural Networks}

\usepackage{listings}
\usepackage{enumitem}
\usepackage{color,colortbl}
\usepackage{soul}
\usepackage{multirow}
\usepackage{amssymb}
\usepackage{amsmath}

\definecolor{Gray}{gray}{0.9}

\begin{document}

\begin{frontmatter}

\title{Private Speech Classification with\\ Secure Multiparty Computation}

\author{Kyle Bittner*} 
\author{Martine De Cock\fnref{myfootnote}}
\address{School of Engineering and Technology\\
University of Washington Tacoma\\
1900 Commerce Street, Tacoma WA 98402-3100, USA\\
\{kyleb29,mdecock\}@uw.edu}
\cortext[mycorrespondingauthor]{Corresponding author}
\fntext[myfootnote]{Guest Professor at Ghent University, Dept.~of Appl.~Math., Comp.~Science and Statistics}

\author{Rafael Dowsley}
\address{Faculty of Information Technology\\
Monash University\\
Australia\\
rafael.dowsley@monash.edu
}

\begin{abstract}
 Deep learning in audio signal processing, such as human voice audio signal classification, is a rich application area of machine learning. Legitimate use cases include voice authentication and emotion recognition. While there are clear advantages to automated human speech classification, application developers can gain knowledge beyond the professed scope from unprotected audio signal processing. In this paper we propose the first privacy-preserving solution for deep learning based audio classification with Secure Multiparty Computation. Our approach allows to classify a speech signal of one party (Alice) with a deep neural network of another party (Bob) without Bob ever seeing Alice's speech signal in an unencrypted manner. As threat models, we consider both passive security, i.e.~with semi-honest parties who follow the instructions of the cryptographic protocols, as well as active security, i.e.~with malicious parties who deviate from the protocols. We evaluate the efficiency-security-accuracy trade-off of the proposed solution in a use case for privacy-preserving emotion detection from speech with a convolutional neural network. In the semi-honest case we can classify a speech signal of 3.5 sec in under 0.3 sec; in the malicious case it takes $\sim$1.6 sec. In both cases there is no leakage of information, and we achieve classification accuracies that are the same as when computations are done on unencrypted data.
\end{abstract}


\end{frontmatter}



\section{Introduction}
Speech technology is becoming increasingly prevalent and intrusive \cite{nautsch2019preserving}. Speech data, i.e.~recordings of human speech, are automatically classified for various purposes, extending from user authentication, to control of services and devices, surveillance, and marketing. The developing prevalence of speech audio processing technology stems from the ever-increasing demand of devices and programs that are ``always-listening'' -- such as smartphones, televisions, and intelligent digital voice assistants -- and the technological improvements in speech technology. Beyond applications that aim to automatically classify speakers or speech, e.g.~for authentication or for emotion detection, respectively, there are countless interesting sound classification tasks\footnote{See e.g.~the IEEE AASP Challenge on Detection and Classification of Acoustic Scenes and Events, http://dcase.community/challenge2020/} that may include speech audio processing. These include gunfire detection in surveillance, cough sensing in healthcare, and noise mitigation enabled by smart acoustic sensor networks \cite{mydlarz2017implementation,salamon2017deep}.

While there are apparent benefits to automated speech audio signal recognition,\footnote{See the Interspeech Computational Paralinguistics Challenges for an overview of applications: \url{http://www.compare.openaudio.eu/}} application developers can gain knowledge beyond the professed scope from unprotected audio signals.
A wealth of personal data can be extracted from speech audio signals, including age and gender, health and emotional state, racial or ethnic origin, geographical background, social identity, and socio-economic status \cite{tomashenkovoiceprivacy}. As stated in the recent survey paper by Nautsch et al. \cite{nautsch2019preserving}, the continued success of speech technologies hinges upon the development of reliable and efficient privacy-preservation capabilities, specifically designed for the automatic processing of speech signals. Efforts to safeguard the privacy of users in data-driven applications are underway along at least three dimensions: (1) by laws and regulations such as the EU General Data Protection Regulation (GDPR) and the California Consumer Privacy Act (CCPA); (2) by anonymization techniques that aim to suppress personally identifiable information in data\footnote{A nice example is the  VoicePrivacy Challenge:\\ \url{https://www.voiceprivacychallenge.org/}}; and (3) by protecting sensitive data through encryption.

\begin{figure}
     \centering
    \includegraphics[width=9cm]{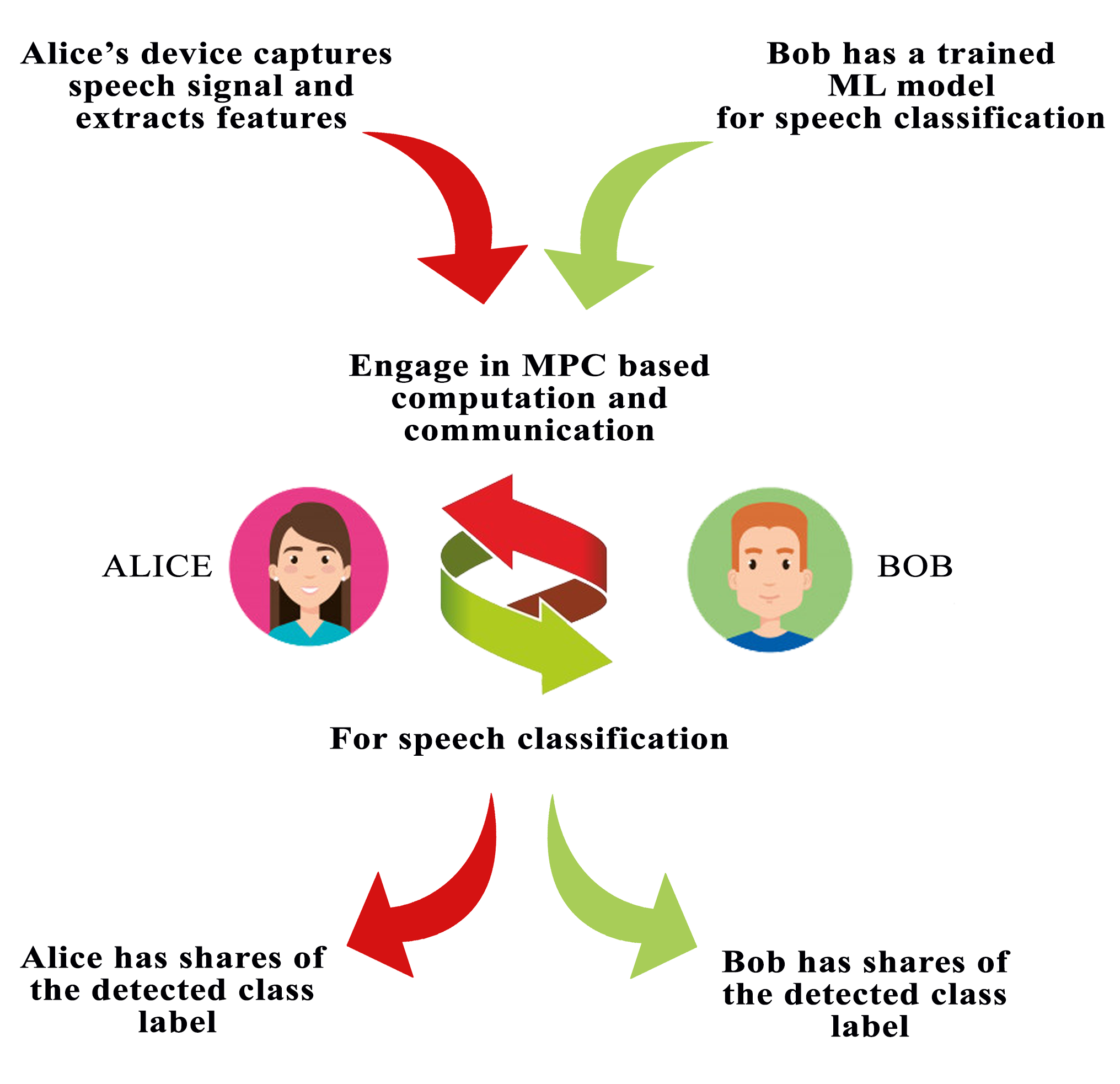}
    \caption{Oblivious speech classification as a two-party computation (2PC) problem in the dishonest majority setting (Section \ref{SEC:ADDITIVESHARING})}
    \label{fig:AliceAndBob}
\end{figure}

In this paper, we focus on the latter, using techniques from Secure Multiparty Computation (MPC). MPC is an umbrella term for cryptographic approaches that allow two or more parties to jointly compute a specified output from their private information in a distributed fashion, without actually revealing the private information to each other \cite{CDN2015}. As illustrated in Figure \ref{fig:AliceAndBob}, speech classification is inherently a two-party computation (2PC) problem, where one party -- nicknamed \textit{Alice} henceforth -- has a speech signal or sound fragment that needs to be classified, and another party -- nicknamed \textit{Bob} -- has a machine learning (ML) classifier that can be used to this end. Similar to how Alice does not want to disclose her speech data to Bob, Bob may not want to disclose his ML model to Alice for a variety of reasons. ML models can be expensive to train and usually constitute a competitive advantage. For example, as reported by Dalskov et al.~\cite{dalskov2019secure}, the network by Yang et al.~\cite{yang2019xlnet} costs between \$61,000
and \$250,000 to train \cite{peng2019}.
Furthermore, deep learning models are powerful enough to memorize specific examples from the training data \cite{carlini2019secret},
hence disclosing a trained model can leak very specific information about the training data, which might be sensitive in itself. Finally, disclosing the trained ML model increases the likelihood that adversaries can develop successful evasion attacks. In the context of speaker or speech characterization, such attacks could consist of altering speech signals to bypass speaker verification systems or to bypass classifiers that detect ``fake speech'', i.e.~that detect the use of speech synthesis tools for malicious purposes such as spreading misinformation, harassment and intimidation \cite{aaai2020fakespeech}.

MPC allows oblivious speech classification through computations over encrypted data. In this way, Alice can classify her speech signal using Bob's model, without Alice revealing her speech signal to anyone in plaintext, and without Bob disclosing his ML model to anyone in-the-clear, i.e.~without encryption. To this end, Alice and Bob engage in computations, and they exchange intermediate encrypted results by communicating with each other. At the end of the oblivious speech classification protocol, Alice and Bob each have ``shares'' of the inferred class label (e.g.~the emotion state of Alice). The true class label is revealed only when these shares are combined, e.g.~when, depending on the application, (1) Bob sends his shares to Alice, or (2) Alice sends his shares to Bob, or (3) both Alice and Bob send their shares to a third party, like a health care provider who might need to be informed when Alice is not doing well.

MPC has already been used for speaker and speech classification with hidden Markov models (HMMs) and Gaussian mixture models (GMMs) \cite{smaragdis2007framework,pathak2013privacy,pathak2012privacy,portelo2014privacy}. While HMMs and GMMs were popular techniques for speech classification in the 1980s and 1990s, more recently deep learning has emerged as a state-of-the-art technique in this field. To the best of our knowledge, MPC-based secure classification of speech with deep neural networks has never been studied. It is this gap in the literature, which is also called out by Nautsch et al.~\cite{nautsch2019preserving}, that we fill in this paper.

Several kinds of neural network architectures can be used for speech classification. As cryptographic methods are known to result in significant increases to computational complexity and/or communication overheads \cite{tomashenkovoiceprivacy}, we choose convolutional neural networks (CNNs), which are computationally less intensive than for instance long short-term memory networks (LSTMs), even without encryption. To the best of our knowledge, all existing work on MPC-based classification with CNNs is developed for and focused on 2-dimensional CNNs, which are commonly used for classification of images. In this paper, we adapt the work that was done in SecureQ8 \cite{dalskov2019secure} for 2-dimensional CNNs (image classification) to 1-dimensional CNNs (speech classification). While speech classification can be done with 2-dimensional inputs (spectrograms), for privacy-preserving speech classification we advocate the use of 1-dimensional CNNs for their smaller model size and efficiency during the secure inference process.

After describing the relationships between this paper and existing work in Section \ref{SEC:RELATED}, in Section \ref{SEC:METHODS} we present details about the proposed methods. These include the pre-processing of the audio and the proposed MPC-friendly neural network architecture, a description of the security settings, and MPC-based protocols for secure classification with 1-dimensional CNNs. We implemented our approach on top of the MP-SPDZ framework \cite{MP-SPDZ}. In Section \ref{SEC:RESULTS}, we present accuracy and runtime results on the RAVDESS benchmark data set \cite{livingstone2018ryerson} for emotion detection from speech. In the active-security setting, i.e.~with malicious adversaries that may deviate from the protocol, a speech signal of 3.5 sec is classified in $\sim$1.6 sec (i.e.~real-time factor 0.46). In the passive-security setting, i.e.~with semi-honest adversaries that adhere to the protocol instructions but try to learn additional information, we can classify a speech signal in 0.26 sec (real-time factor 0.07). The accuracy in both cases is 70.32\%, which is in the range of state-of-the-art approaches for emotion recognition from the RAVDESS dataset in-the-clear.
Furthermore, our approach is provably secure: during the secure inference, nobody other than Alice learns anything about her speech signal, and nobody other than Bob learns anything about his model parameters. Obviously, disclosing the inferred class label at the end of the protocol to Bob reveals information about Alice.
Similarly, disclosing class labels to Alice could reveal some information about Bob's model, which could allow membership inference and/or model inversion attacks. To protect against these, Differential Privacy (DP) could be used to add controlled noise to the gradients during DNN training, which would lead to a decrease in accuracy. We consider this to be orthogonal to, and outside of, the scope of this paper. 

As we highlight in Section \ref{SEC:CONCLUSION},
our results answer a question that has remained open in the literature thus far, namely to what extent MPC-based protocols can enable provably secure and highly accurate real-time speech classification \cite{nautsch2019preserving}.



\section{Related work}\label{SEC:RELATED}
We refer to the work of Nautsch et al.~\cite{nautsch2019preserving} for an excellent and comprehensive survey of existing work on privacy-preserving speaker and speech characterization. Below we focus on what is most relevant for our work, namely (1) existing approaches to speech classification that are based on MPC, and (2) existing work on secure inference with a trained deep learning model based on MPC, for applications other than speech or speaker characterization. To the best of our knowledge, none of the existing work in category (1) is based on deep learning, while none of the existing work in (2) has been applied to 1-dimensional CNNs in general, or to speech classification in specific. This is the gap we close in our work. We note that, prior to our work, an MPC based approach for speech classification based on neural networks has been considered highly impractical due to an assumed massive overhead in computation time and communication costs \cite{brasser2018voiceguard}. This has prompted research into the use of trusted execution environments, such as real-time speech classification based
on Intel SGX \cite{brasser2018voiceguard}, and real-time keyword
recognition on ARM-based (mobile) devices using TrustZone capabilities \cite{bayerl2020offline}.
Contrary to this, the solution we present in this paper does not require special hardware, and, as we demonstrate in Section \ref{SEC:RESULTS}, is fast enough for use in real-time. 

\textbf{(1) MPC-based speech classification.} In the clear, i.e.~without concern for user privacy, there are several successful ML approaches for speech classification. Well-known ML work horses that gained popularity in the 1980s and 1990s are hidden Markov models (HMMs) and Gaussian mixture models (GMMs). The earliest work on privacy-preserving speech classification based on MPC focused on the design of cryptographic protocols to make training and inference with HMMs and GMMs secure in the semi-honest setting \cite{smaragdis2007framework,pathak2013privacy,pathak2012privacy}. These early approaches were based on homomorphic encryption (HE) and slow because of the large computation costs. Port\^{e}lo et al. \cite{portelo2014privacy}~substantially improved upon this computational cost by using Garbled Circuits (GC) instead of HE, in a GMM based protocol specifically for speaker verification, i.e.~voice based authentication.
Similarly, Treiber et al. \cite{treiber2019privacy}~introduced an MPC approach for privacy-preserving speaker verification based on secure computation of the cosine of a vector with biometric characteristics extracted from the speaker's audio signal on one hand, and a stored reference embedding on the other hand; we note that the comparison of two vectors for \textit{speaker verification} is computationally a less involved task than inference with a trained ML model for \textit{speech classification}, as we do in this paper.


While up until a decade ago, HMM used to be popular for speech processing and audio classification, more recently deep learning has been acknowledged as a state-of-the-art ML approach in this field \cite{trigeorgis2016adieu,milde2015using}. The CNN approach that we follow in this paper adheres to the latter.

\textbf{(2) MPC-based classification with CNNs.} The problem of doing privacy-preserving inference with trained neural networks has received a lot of attention in the literature recently, and a variety of MPC-based approaches and frameworks have been proposed. Most of these, including MiniONN \cite{liu2017oblivious}, SecureML \cite{mohassel2017secureml}, DeepSecure \cite{rouhani2018deepsecure}, Chameleon \cite{riazi2018chameleon}, Gazelle \cite{juvekar2018gazelle}, Quotient \cite{agrawal2019quotient}, XONN \cite{riazi2019xonn}, and Delphi \cite{mishra2020delphi}, are limited to the semi-honest security setting, i.e.~they guarantee that no information is leaked as long as the parties honestly execute the protocols. CrypTFlow \cite{kumar2020cryptflow}, based in part on SecureNN \cite{wagh2019securenn}, is an interesting recent addition to the growing body of MPC-based secure inference frameworks.   
In addition to the semi-honest case, CrypTFlow also guarantees security in the malicious case, where parties may deviate arbitrarily from the protocols. To this end, CrypTFlow uses a combination of cryptographic techniques and secure hardware (Intel SGX). To the best of our knowledge, SecureQ8 \cite{dalskov2019secure} is the only work so far on MPC-based secure inference with trained CNNs in both the semi-honest and malicious case that does not require special secure hardware. In this paper, we adapt the work that was done in SecureQ8 for 2-dimensional CNNs (image classification) to 1-dimensional CNNs (speech classification).

Outside of MPC, we mention the research by Dias et al.~\cite{dias2018exploring} and Teixeira et al.~\cite{teixeira2019privacy} who combine neural networks with (leveled) fully homomorphic encryption (FHE) for privacy-preserving detection of emotion and of voice-affecting diseases such as a cold, a depression, and Parkinson's disease. The main difference between their work, which builds on Cryptonets \cite{gilad2016cryptonets}, and ours, is that in \cite{dias2018exploring,teixeira2019privacy}, Alice encrypts her input feature vector and sends it to Bob, who uses FHE to perform computations over the encrypted data, while in our MPC approach both Alice and Bob perform computations. 
FHE comes with lower communication costs than MPC, at the expense of substantially higher computation costs, which could make it prohibitively expensive for audio classification. Dias et al.~\cite{dias2018exploring} use a relatively simple architecture, namely a multi-layer perceptron (MLP) with two hidden layers, and no convolutional layers. Based on runtime results reported in \cite{gilad2016cryptonets}, the CNN approach that we propose in this paper, is faster than the MLP approach of Dias et al.~\cite{dias2018exploring}, even for active security (i.e.~malicious parties), and an estimated two orders of magnitude faster for passive security (i.e.~semi-honest parties).



\section{Methods}\label{SEC:METHODS}
Our approach for private speech characterization consists of two phases: first the ML model training is done by Bob in the clear, i.e.~on training data that is not encrypted (see Section \ref{SEC:MODELTRAIN}), then the inference with the trained model is performed securely using a MPC-based solution
(see Section \ref{SEC:INFERENCE}). In the secure inference steps, all computations are done over encrypted data and model parameters, meaning that Alice does not learn anything about Bob's model weights or training examples, while Bob does not learn anything about Alice's speech signal.

\subsection{Data preprocessing and neural network architecture}\label{SEC:MODELTRAIN}

\subsubsection{Data preprocessing and feature extraction} 
\label{SEC:DATAPREPROC}
Our assumption is that Bob has a set of audio files (speech signals) that are each annotated with a label, and he uses these to train an ML model that can assign a correct label to a previously unseen audio file (Alice's input). It is common in speech processing for classifiers to work on features extracted from the speech signal as opposed to on the raw speech signal itself. These features and the software to extract them are widely known and publicly available. It is for example very common to convert a speech signal into a sequence of feature vectors of mel-frequency cepstral  coefficients (MFCC) \cite{davis1980comparison} that are extracted from sliding windows of consecutive speech.
We assume that Bob converts each audio file from his training data into a sequence of $r$ feature vectors, each of length $m$, and subsequently averages them to obtain one feature vector of length $m$ per audio file. Similarly, Alice converts her speech signal into a sequence of $r$ vectors of MFCC coefficients, averages them, and uses the resulting feature vector of length $m$ as her input to the protocol for speech classification (see Figure \ref{fig:AliceAndBob}).
As we demonstrate in Section \ref{SEC:RESULTS}, we can train highly accurate ML models for speech classification based on these extracted feature vectors. That in itself is clear evidence that the feature vectors contain meaningful, private information that needs to be kept private during inference, as we do with the technique described in Section \ref{SEC:INFERENCE}.

\subsubsection{MPC-friendly CNN model architecture}
\label{SEC:CNNARCH}
We propose the use of a standard, MPC-friendly CNN model architecture. By ``MPC-friendly'' we mean that the operations to be performed when doing inference with the trained CNN are chosen purposefully among operations for which efficient MPC protocols exist. 
A standard CNN contains one or more blocks that each have a convolutional layer, followed by an activation layer, and an optional pooling layer. 

The difference between the more commonly used 2-dimensional CNNs on one hand, and the 1-dimensional CNNs that we use in this paper, is that in a 1-dimensional CNN, convolutional operations are performed across one dimension. In 2-dimensional CNNs, the shape of the input of a convolutional layer is defined in terms of its height H, width M, and depth D. In a 1-dimensional CNN, the height is always 1, hence the input $X$ is a D $\times$ M matrix, as illustrated in Figure \ref{FIG:1DCONV}. In this paper, the input to the first convolutional layer in the CNN is a vector of $m$ MFCC values as explained in Section \ref{SEC:DATAPREPROC}, hence D$=1$ and M$=m$. In further convolutional layers, the depth D is typically larger, and determined by the number of filters (kernels) in a previous layer.
A convolutional layer is defined by F filters, each of size D $\times$ L, with L the width of the filters. In addition, for each filter $W_i$, $i=1,\ldots, \mbox{F}$, the convolutional layer contains a bias term $b_i \in \mathbb{R}$. The values of the weights in the filters and the bias are, as usual, learned during training. The output produced by a convolutional layer with F filters $W_i$ of width L, when applied to an input $X$ of size D $\times$ M, is a matrix $Z$ of size F $\times$ M, which is computed as:\footnote{For example, in the CNN in Figure} \ref{FIG:ARCHITECTURE}, in the first convolutional layer (line (2) in the code), D$=1$, M$=40$, F$=128$ and L$=5$. The input to this layer is a $1 \times 40$ matrix, and the output of the layer is a $128 \times 40$ matrix, which is downsampled by the average pooling layer on line (5) to a $128 \times 10$ matrix. In the second convolutional layer (line (6) in the code), D$=128$, M$=10$, F$=128$ and L$=5$. The output of this layer is a $128 \times 10$ matrix.
\begin{equation}\label{EQ:CODEFORLOOPS}
\begin{tabular}{l}
\mbox{for\ } $i \leftarrow 1$ \mbox{to\ F}\\
\phantom{xxxx} \mbox{for\ } $j \leftarrow 1$ \mbox{to\ M}\\
\phantom{xxxxxxxx} $Z_{i,j} \leftarrow X[j:j+\mbox{L}-1] \odot W_i + b_i$\\
\end{tabular}
\end{equation}

\noindent
In this pseudocode, $X[j:j+\mbox{L}-1]$ denotes the submatrix of $X$ that consists of column $j$ through column $j+\mbox{L}-1$ of $X$,\footnote{We assume a stride of 1, and zero padding, which means that $\mbox{L}-1$ columns with 0s are appended to $X$ to avoid the index from running out of bounds.} while $\odot$ denotes the Frobenius inner product (a generalization to matrices of the dot product of vectors). The computation of the $i$th row of $Z$ is illustrated in Figure \ref{FIG:1DCONV}. In privacy-preserving speech classification, the input $X$ into the first convolutional layer is known to Alice, while the values of $W_i$ and $b_i$ are known to Bob. We address in Section \ref{SEC:INFERENCE} how in this case $Z$  can be computed, and subsequent CNN operations can be performed, without the need for Alice to disclose $X$ and without the need for Bob to disclose $W_i$ and $b_i$.

\newcolumntype{g}{>{\columncolor{Gray}}c}
\begin{figure}
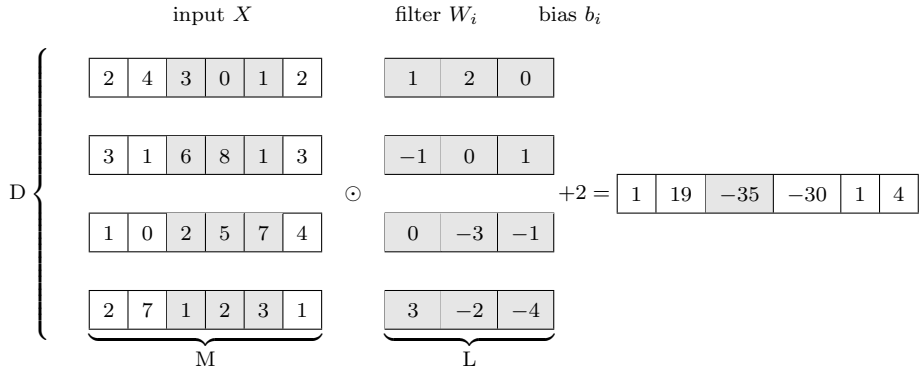

\footnotesize{
$
\ \ \ \ \ \ \ \ \ \ \ \ \ \ \ \ \ \ \ \ \ \ \mbox{input\ } X 
\ \ \ \ \ \ \ \ \  
\ \ \ \ \ \ \ \ \ \ \mbox{filter\ }W_i 
\ \ \ \ \ \ \ \ \mbox{bias\ }b_i
$
$$
\mbox{D}
\left\{
\begin{tabular}{c}
\\
\\
\\
\\
\\
\\
\\
\\
\end{tabular}
\right.
\underbrace{
\begin{array}{|c|c|g|g|g|c|}
\hline
2 & 4 & 3 & 0 & 1 & 2\\
\hline
\multicolumn{6}{c}{}\\
\hline
3 & 1 & 6 & 8 & 1 & 3\\
\hline
\multicolumn{6}{c}{}\\
\hline
1 & 0 & 2 & 5 & 7 & 4\\
\hline
\multicolumn{6}{c}{}\\
\hline
2 & 7 & 1 & 2 & 3 & 1\\
\hline
\end{array}
}_{\mbox{M}}
\, \, \, \, \,
\odot
\, \, \, \, \,
\underbrace{
\begin{array}{|g|g|g|}
\hline
1 & 2 & 0\\
\hline
\multicolumn{3}{c}{}\\
\hline
-1 & 0 & 1\\
\hline
\multicolumn{3}{c}{}\\
\hline
0 & -3 & -1\\
\hline
\multicolumn{3}{c}{}\\
\hline
3 & -2 & -4\\
\hline
\end{array}
}_{\mbox{L}}
+
2
= 
\begin{array}{|c|c|g|c|c|c|}
\hline
1 & 19 & -35 & -30 & 1 & 4\\
\hline
\end{array}
$$
}
\caption{Illustration of a 1-dimensional convolution on an input width M=6, height H=1, and depth D=4. This is equivalent to a 2-dimensional convolution on an input of width 6, height 4, and depth 1. The filter ``slides'' from the left to the right. \label{FIG:1DCONV}}
\end{figure}

As the activation function in the convolutional blocks, we choose the RELU function $f(z) = \max(0,z)$, which means that all negative values are mapped to $0$. For the pooling layer, we select average pooling instead of max pooling, because in an MPC setting additions and division by a publicly known constant (as is needed to compute an average) are computationally less expensive than performing comparisons (which would be needed to find a maximum). Applying RELU and average pooling with size 2 to the output in Figure \ref{FIG:1DCONV} would yield $[10,0,2.5]$.

The stacked convolutional blocks are followed by a dense layer, the application of which comes down to a product of two matrices. The activation function on the final layer is typically a logistic function (for binary classification problems) or a softmax operation (for multi-class classification problems). The output of the softmax operation is a probability for each of the possible class labels; the label with the highest probability is returned as the final result. While the use of a softmax function is important during training, we note that during inference it can be replaced by an argmax function. Indeed, the softmax operation does not change the ordering among the logits, i.e.~the values that are passed into it from the previous layer. Argmax is computationally much less expensive to compute in a privacy-preserving manner.
Finally, any dropout layers that are used to improve the training process, are omitted during inference, which means that we do not need to include MPC-based protocols for these layers when doing secure inference (see Section \ref{SEC:INFERENCE}). We refer to Section \ref{SEC:RESULTS} for more details about the exact CNN architecture that we used in our experiments.

\subsection{Privacy-preserving inference with a 1-dimensional CNN}\label{SEC:INFERENCE}

After giving a high level overview in Section \ref{SEC:SETTINGS} of the security settings that we consider, we recall the principles of MPC based on secret sharing in Section \ref{SEC:ADDITIVESHARING} and \ref{SEC:REPLICATEDSHARING}, and the particular MPC schemes that we use in this paper. This includes an explanation of how Alice and Bob can perform additions and multiplications on integers even if they only have so-called shares of the integers instead of the actual values. Since in speech classification, Alice's MFCC feature values and Bob's model parameters are real numbers, in Section \ref{SEC:QUANT} we explain how Alice and Bob use techniques from quantization of neural networks to convert their floating-point data into integers before they execute the MPC protocols. Next we explain in Section \ref{SEC:SECCLASSIFICATION} how Alice and Bob can use an MPC scheme to perform privacy-preserving speech classification. 

\subsubsection{Security settings}\label{SEC:SETTINGS} 
There exist a variety of MPC schemes, 
designed for different numbers of participants and offering various levels of security that correspond to different threat models. In the scenario of privacy-preserving speech classification that we consider in this work, there are two participants, Alice and Bob, and one of them may be corrupted.
When Alice and Bob execute a secure MPC protocol between themselves to perform the privacy-preserving speech classification, as illustrated in Figure \ref{fig:AliceAndBob}, one corrupted party means that we are in the so-called scenario of \textit{dishonest majority}. In general, a dishonest majority setting is one where an adversary can corrupt a fraction of the protocol participants that is equal to or greater than 1/2. 
In our two-party computation (2PC) setting this means that each party can only trust itself and assumes that the other party may be corrupted. We describe the MPC protocols that we use for the dishonest-majority setting in Section \ref{SEC:ADDITIVESHARING}.

\begin{figure}
     \centering
    \includegraphics[width=9cm]{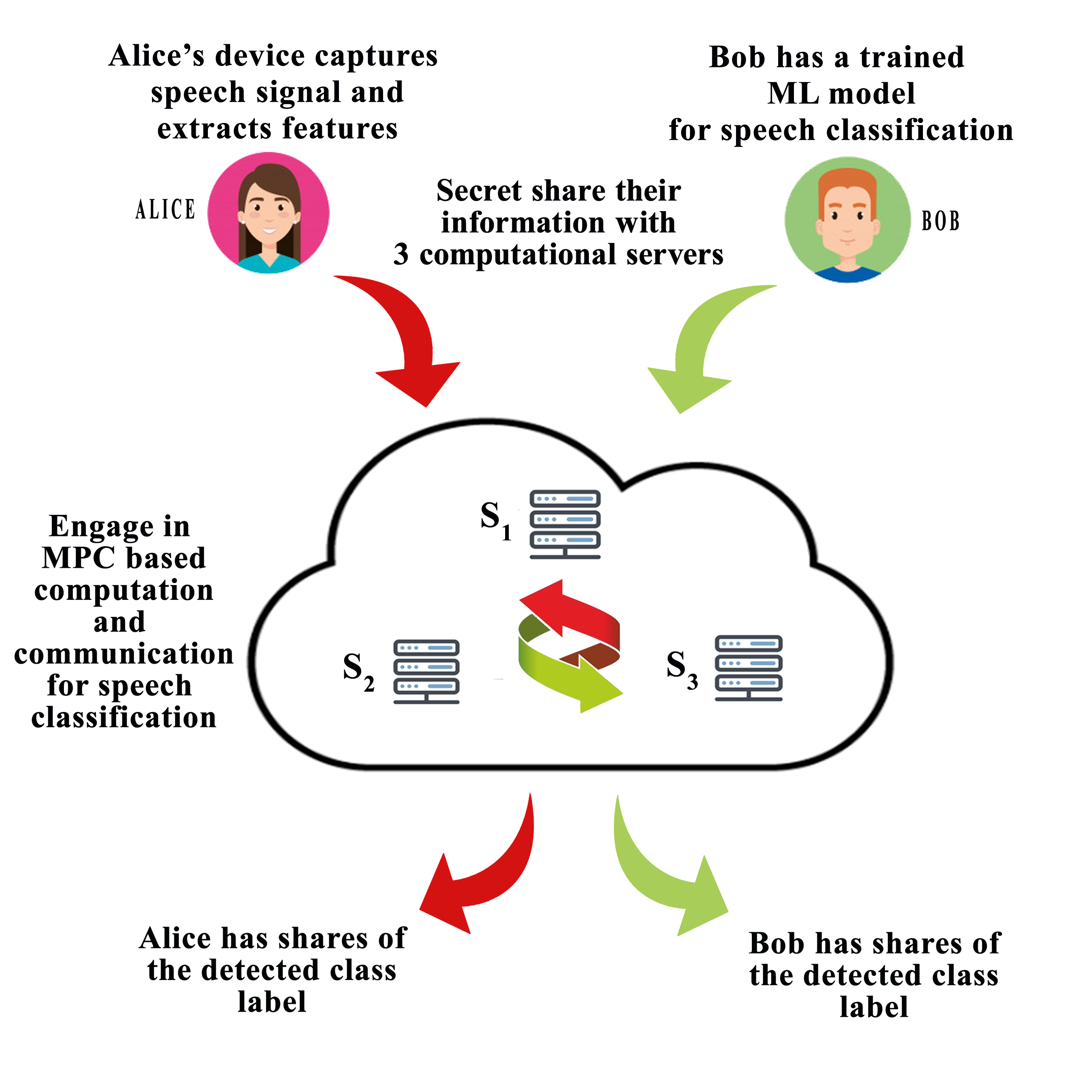}
    \caption{Oblivious speech classification as a three-party computation (3PC) problem in the honest-majority setting (Section \ref{SEC:REPLICATEDSHARING}). Alice and Bob outsource the computations to 3 servers.}
    \label{fig:AliceAndBob3PC}
\end{figure}

MPC protocols in the dishonest-majority setting such as the 2PC scenario from Figure \ref{fig:AliceAndBob} are much more computationally expensive than protocols in an \textit{honest-majority} setting, i.e.~when more than half of the protocol participants are honest. Therefore, many works on privacy-preserving inference have considered the setting in which Alice and Bob outsource the secure computations to a set of 3 or more servers, of which a majority is assumed to be honest (e.g., \cite{dalskov2019secure,riazi2018chameleon,kumar2020cryptflow,wagh2019securenn}). In this work we also evaluate the performance of privacy-preserving 
speech classification in the scenario in which Alice and Bob outsource the secure classification to 3 servers (three-party computation, 3PC), one of which can be corrupted. The protocols that we use for this scenario, which is illustrated in Figure \ref{fig:AliceAndBob3PC}, are described in Section \ref{SEC:REPLICATEDSHARING}.

Furthermore, a party can be corrupted in different ways. In the \textit{passive-security} setting (also known as \textit{semi-honest} or \textit{honest-but-curious adversaries}), the corrupted parties follow the specified protocol instructions, but they may try to learn additional information (i.e., information other than what can be inferred from their specified inputs and outputs)
from the messages exchanged during the protocol execution. Secure MPC protocols prevent such information leakage. In the \textit{active-security} setting (also known as \textit{malicious adversaries}), the parties may deviate from the protocol instructions in arbitrary ways, for instance by providing incorrect values on purpose. In this case, secure MPC protocols should prevent information leakage and detect devious behavior. Protection against such a stronger threat model comes at a higher computational cost. 

In this paper, we evaluate multiple MPC schemes and their efficiency-security-accuracy trade-off for privacy-preserving speech classification.

\subsubsection{Secret sharing-based MPC for dishonest majority}\label{SEC:ADDITIVESHARING} 
In the MPC schemes that we use, all computations are done on integers, modulo an integer $q$.  The modulo $q$ is a hyperparameter that defines the algebraic structure in which the computations are done, which in turn has a direct effect on the efficiency of the MPC protocols for different tasks. In Section \ref{SEC:RESULTS}, we evaluate MPC schemes where $q$ is a prime number as well as where $q$ is a power of 2.

Furthermore, all MPC schemes for the dishonest-majority scenario that we use are based on \textit{additive sharing}. A value $x$ in $\mathbb{Z}_q =\{0,1,\ldots,q-1\}$ is secret shared between Alice and Bob by picking uniformly random values $x_1, x_2 \in \mathbb{Z}_q$ such that 
\begin{equation}
    x_1 + x_2  =  x \mod{q}. \label{EQ:ADDSHARES}\\
\end{equation}
Equation (\ref{EQ:ADDSHARES}) expresses that $x_1$ and $x_2$ are additive shares of $x$ (which are delivered to Alice and Bob, respectively). Note that no information about the secret value $x$ is revealed by any of the individual shares $x_1$ or $x_2$, but the secret-shared value can be trivially revealed by combining both shares. As we explain below, the parties Alice and Bob can jointly perform computations on numbers by performing computations on their own shares, without the parties learning the values of the numbers themselves.\footnote{We often omit the modular notation for conciseness.}

\textbf{Passive security.} For protocols in the passive-security setting, we use $[\![x]\!]$ as a shorthand for a secret sharing of $x$, i.e. $[\![x]\!] = (x_1,x_2)$. Given secret-shared values $[\![x]\!] = (x_1,x_2)$ and $[\![y]\!] = (y_1,y_2)$, and a constant $c$, Alice and Bob can jointly perform the following operations, each by doing only local computations on their own shares:
\begin{itemize}[leftmargin=*]
  \item Addition of a constant ($z=x+c$): Alice and Bob compute 
  $(x_1+c,x_2)$. Note that Alice adds $c$ to her share $x_1$, while Bob keeps the same share $x_2$. This operation is denoted by $[\![z]\!]\leftarrow[\![x]\!] + c$.
 
  \item Addition ($z=x+y$): Alice and Bob compute
  $(x_1+y_1,x_2+y_2)$ by adding their local shares of $x$ and $y$. This operation is denoted by 
  $[\![z]\!] \leftarrow[\![x]\!]+[\![y]\!]$. 
  
  \item Multiplication by a constant ($z=c \cdot x$):
   Alice and Bob compute $(c \cdot x_1, c \cdot x_2)$
   by multiplying their local shares of $x$ by $c$. This operation is denoted by 
  $[\![z]\!]\leftarrow c[\![x]\!]$. 
\end{itemize}

The basic operation that is missing in the list above is the multiplication of secret-shared values $[\![x]\!]$ and $[\![y]\!]$. This is done using a so-called \textit{multiplication triple} \cite{Beavertriple}, which is a triple of secret-shared values $[\![a]\!]$, $[\![b]\!]$, $[\![c]\!]$, such that $a$ and $b$ are uniformly random values in $\mathbb{Z}_q$ and $c=a \cdot b$.
We explain later how Alice and Bob obtain such multiplication triples. Given that they have a multiplication triple, Alice and Bob can compute
$[\![d]\!] = [\![x]\!] - [\![a]\!]$ and $[\![e]\!] = [\![y]\!] - [\![b]\!]$, and, in a communication step, \textit{open} $d$ and $e$ by disclosing their respective shares of $d$ and $e$ to each other. Next, they can compute
$[\![z]\!] = [\![c]\!] + d \cdot [\![b]\!] + e \cdot [\![b]\!] + d \cdot e$, which is equal to $[\![x \cdot y]\!]$. Each multiplication requires a fresh multiplication triple; generating sufficient multiplication triples contributes substantially to the computational cost of MPC protocols. This secure multiplication protocol can be generalized for the multiplication of element-wise secret-shared matrices (for efficiency gains) using matrix multiplication triples, and keeps its security even when composed with other arbitrary building blocks \cite{IEEETDSC:CDHK+17,Dowsley16}.

\textbf{Active security.}
In the case of active security, the main idea to prevent the players from cheating is to use a Message Authentication Code (MAC). We focus first on the case of a prime field $\mathbb{Z}_q$, with $q$ a prime number. To verify the correctness of the computations, Alice and Bob each have a share of a fixed MAC key $\alpha \in \mathbb{Z}_q$, i.e.~Alice has $\alpha_1$ and Bob has $\alpha_2$ such that $\alpha_1 + \alpha_2 = \alpha \mod q$. When a value $x$ is secret shared between Alice and Bob, they also get shares $m_1$ and $m_2$, respectively, of a MAC such that 
\begin{equation}
m_1 + m_2  =  x \cdot \alpha \mod{q} \label{EQ:MACREL}
\end{equation}

Equation (\ref{EQ:MACREL}) is the so-called MAC relation. If at any point Alice and Bob need to \textit{open} a secret-shared number, i.e.~make its value $x$ known, then Alice first reveals $x_1$ to Bob, while Bob reveals $x_2$ to Alice, so that they both can compute $x=x_1+x_2$. Next, to verify that the MAC relation holds, Alice commits the value of $m_1-x\cdot \alpha_1$ while Bob commits the value of 
$m_2-x\cdot \alpha_2$, and they subsequently reveal these values to each other so that they can both verify that they add up to 0. The purpose of the commit phase before the reveal phase is to prevent the parties from changing their value based on what the other party reveals.\footnote{In practice the verification of all MAC relations is performed in an aggregated, much more efficient way right before the end of the protocol.} 

We continue to use $[\![x]\!]$ as a shorthand for a secret sharing of $x$ in the case of active security, but in this case $[\![x]\!] = (x_1,x_2,m_1,m_2)$. Given secret-shared values $[\![x]\!] = (x_1,x_2,m_1,m_2)$ and $[\![y]\!] = (y_1,y_2,n_1,n_2)$, and a constant $c$, Alice and Bob can jointly perform the same operations as before, with only local computations:
\begin{itemize}[leftmargin=*]
  \item Addition of a constant ($z=x+c$): Alice and Bob compute 
  $(x_1+c,x_2,\alpha_1 \cdot c + m_1, \alpha_2 \cdot c + m_2)$. Note that the MAC relation remains satisfied, since
  $\alpha_1 \cdot c + m_1 + \alpha_2 \cdot c + m_2 = (\alpha_1 + \alpha_2) \cdot c + (m_1 + m_2) = \alpha \cdot c + x \cdot \alpha = \alpha \cdot (x+c)$.
 
  \item Addition ($z=x+y$): Alice and Bob compute
  $(x_1+y_1,x_2+y_2,m_1+n_1,m_2+n_2)$ by adding their local shares of $x$ and $y$. The MAC relation remains trivially satisfied.
  
  \item Multiplication by a constant ($z=c \cdot x$):
   Alice and Bob 
   compute
   $(c \cdot x_1, c \cdot x_2, c \cdot m_1, c \cdot n_2)$
   by multiplying their local shares of $x$ by $c$. The MAC relation remains trivially satisfied.
\end{itemize}

The notations we use for the operations are the same as in the passive-security case. In the case of the protocol with active security using binary fields $\mathbb{Z}_q$, with $q$ a power of 2, there are a few additional technical details regarding the MAC, but the MPC scheme provides the same set of basic local operations that are described above. We refer interested readers to \cite{cramer2018spd} for further details.

In the case of active security, the multiplication of secret-shared values can also be performed as described above using multiplication triples, but the multiplication triples must be generated together with the respective MACs. 

Additionally, since in the case of active security all secret-shared values $x$ that are used in the computations must contain a corresponding MAC (defined by $m_1$ and $m_2$ as in Equation \ref{EQ:MACREL}), a procedure for the parties to obtain a MAC for their inputs must be used. This is done as follows. During the offline phase  described below, a secret sharing $[\![r]\!]$ (with a MAC) of a random value $r \in \mathbb{Z}_q$ is generated and distributed to Alice and Bob. If Alice has an input $a \in \mathbb{Z}_q$, the secret sharing $[\![r]\!]$ is opened towards her and she sends $c=a-r$ to Bob. They then compute the secret sharing $[\![a]\!]\leftarrow[\![r]\!] + c$, which contains a MAC. Note that the value $c$ is uniformly random and independent from $a$, and therefore does not reveal any information to Bob.

\textbf{Generation of Multiplication Triples and Random Values During the Offline Phase.} For performance reasons, modern MPC schemes are normally divided in two phases: the offline and online phases. The offline phase only performs computations that are independent from the specific inputs of the parties to the protocol (Alice's speech signal and Bob's trained model parameters), and therefore can be executed far before the inputs are fixed. Modern MPC protocols try to perform as much of the computation as possible in the off\-line phase, so that the online phase can be faster, improving the responsiveness of the MPC solution. 

In the case of the secret-shared based schemes that we consider, the computationally heavy operations are the generation of the multiplication triples and of the random values, and both of them are independent of the specific inputs of parties and can be delegated to the offline phase, whose main purpose is to generate these values. The parties can jointly generate them using techniques such as homomorphic encryption or oblivious transfer. During the online phase the parties only need to perform basic arithmetic operations, whose computational costs are quite small.

\textbf{MPC Schemes.} Table \ref{TAB:MPCSCHEMES} contains an overview of the MPC schemes that we use in this paper. The MPC schemes for passive security provide protection against semi-honest adversaries, while the MPC schemes for active security provide protection against malicious adversaries. The distinction between the underlying algebraic structures $\mathbb{Z}_p$ and $\mathbb{Z}_{2^k}$ is meaningful because of its potential impact on the efficiency of the protocols. We briefly describe each MPC scheme for the dishonest majority scenario here (the ones for the honest majority scenario are described in Section \ref{SEC:REPLICATEDSHARING}):

\begin{table}
\centering
\begin{tabular}{c|cc}
dishonest majority & $\mathbb{Z}_p$ & $\mathbb{Z}_{2^k}$\\
\hline
passive security & SEMI & SEMI2K \\
active security  & MASCOT & SPDZ2K\\
\multicolumn{3}{c}{}\\
honest majority & $\mathbb{Z}_p$ & $\mathbb{Z}_{2^k}$\\
\hline
passive security & ReplPrime & Repl2k \\
active security  & PsReplPrime & PsRepl2k\\
\end{tabular}
\caption{Overview of MPC schemes according to threat model and algebraic structure \label{TAB:MPCSCHEMES}}
\end{table}

\begin{itemize}[leftmargin=*]
\item 
In the case of active security using a prime field, we use \textbf{MASCOT} \cite{keller2016mascot}, an MPC scheme with an improved offline phase based on oblivious transfer techniques to generate the necessary values for the online phase of the SPDZ protocol \cite{damgaard2012multiparty} (which is the online phase described above). Note that the offline phase is also performed between Alice and Bob, and one of them may act maliciously. Therefore it is necessary to use a series of mechanisms (such as consistency checking, privacy amplification techniques, and oblivious transfer checks) in order to guarantee that correct multiplication triples and random values are generated and that nothing about them leaks to Alice or Bob. We point interested readers to \cite{keller2016mascot} for further details about how these values are generated in the offline phase.  

\item
In the case of active security using a binary field, we use \textbf{SPDZ2K} \cite{cramer2018spd}. It adapts the offline phase of \textbf{MASCOT} to generate multiplication triples and random values for a binary  field, which are then consumed by its online phase (which is an adaptation of SPDZ to the setting of binary fields). See \cite{cramer2018spd} for further details. 

\item
For passive security we use \textbf{SEMI} for the case of prime fields, and \textbf{SEMI2K} for binary fields. Both schemes generate multiplication triples using techniques based on oblivious transfer. \textbf{SEMI} is a cut-down version of \textbf{MASCOT}, which eliminates all the additional machinery of \textbf{MASCOT} that is only necessary for the case of active security (such as consistency checking, privacy amplification techniques, the generation and use of message authentication codes, and oblivious transfer checks). Similarly, \textbf{SEMI2K} is a cut-down version of \textbf{SPDZ2K} to focus on passive security. 
\end{itemize}

Many previous works on privacy-preserving machine learning have assumed the existence of a trusted initializer (e.g., \cite{david2015efficient,AISec:CDNN15,fritchman2018,IEEENSRE:ADMW+19,NeurIPS2019,guo2020secure,idash}), who pre-distributes correlated randomness to the protocol participants at a setup phase and does not participate in any other part of the protocol execution. Note that such a trusted initializer would completely eliminate the need of executing the offline phase of the above protocols, as the trusted initializer can pre-distribute all necessary multiplication triples and random values to Alice and Bob. However, we are interested in evaluating the performance of secure classification in the setting in which \textit{no such trusted initializer is available} to the model and data owners and they have to execute the complete two-party computation solution between themselves.

\subsubsection{Secret sharing-based MPC for honest majority}\label{SEC:REPLICATEDSHARING} 

In the setting with 3 computing servers and at most 1 corruption (i.e., honest majority setting), we use MPC schemes based on replicated secret sharing, which allow much faster solutions than in the two-party setting. 

In a replicated secret sharing scheme, a value $x$ in $\mathbb{Z}_q =\{0,1,\ldots,q-1\}$ is secret shared among servers $S_1, S_2$ and $S_3$ by picking uniformly random values $x_1, x_2, x_3 \in \mathbb{Z}_q$ such that 
\begin{equation}
    x_1 + x_2 +x_3 =  x \mod{q}, \label{EQ:REPSHARES}\\
\end{equation}
and distributing $(x_1,x_2)$ to $S_1$, $(x_2,x_3)$ to $S_2$, and $(x_3,x_1)$ to $S_3$. Note that no single server can obtain any information about $x$ given its share. We continue to use $[\![x]\!]$ as a shorthand for a secret sharing of $x$ in this case.

\textbf{Passive security.} As in the case of additive secret sharings, the 3 parties can easily perform the following operations through carrying out local computations: addition of a constant, addition of secret-shared values, and multiplication by a constant. The biggest advantage of this replicated secret sharing scheme is that it enables a more efficient procedure for multiplying secret-shared values. When multiplying $x \cdot y=(x_1 + x_2 +x_3)(y_1 + y_2 +y_3)$, the servers can 
locally perform the following computations: $S_1$ computes $z_1=x_1 \cdot y_1+x_1 \cdot y_2+x_2 \cdot y_1$, $S_2$ computes $z_2=x_2 \cdot y_2+x_2 \cdot y_3+x_3 \cdot y_2$ and $S_3$ computes $z_3=x_3 \cdot y_3+x_3 \cdot y_1+x_1 \cdot y_3$. After performing these local computations, the servers obtain an additive secret sharing of $x \cdot y$ without needing any interactions. Next, they just need to convert from the additive secret sharing representation back to a replicated secret sharing representation, so that it is possible to perform more multiplications in the same way. In order to securely do this conversion, the servers obtain an additive secret sharing of $0$ by picking uniformly random $u_1,u_2,u_3$ such that $u_1 + u_2 +u_3 = 0$, which can be locally done with computational security by using pseudorandom functions, and $S_i$ locally computes $v_i=z_i+u_i$. Finally, $S_1$ sends $v_1$ to $S_3$, $S_2$ sends $v_2$ to $S_1$, and $S_3$ sends $v_3$ to $S_2$, enabling the servers $S_1, S_2$ and $S_3$ to get the replicated secret shares $(v_1,v_2)$, $(v_2,v_3)$, and $(v_3,v_1)$, respectively, of the value $v=x \cdot y$. Note that for performing the multiplication of secret-shared values, each server only needs to send a single ring element to one other server, and no expensive public-key encryption operations (such as homomorphic encryption or oblivious transfer)
are required. This MPC scheme was introduced by Araki et al. \cite{araki2016high}; we refer to the original paper for further details. Referring back to the second part of Table \ref{TAB:MPCSCHEMES}, we denote the version working on a prime field by \textbf{ReplPrime}, and the version working on a binary field by \textbf{Repl2k}.

\textbf{Active security.} In the case of malicious adversaries, the MPC scheme \textbf{PsReplPrime} that we consider for prime fields uses the approach introduced by Lindell and Nof \cite{lindell2017framework} of generating multiplication triples optimistically in the offline phase (i.e., running the multiplication protocol that is secure against semi-honest adversaries), performing the triple verification via sacrificing
(i.e., one additional triple is used to verify the triple in question and this additional triple is then discarded \cite{lindell2017framework}), and then using Beaver's protocol for multiplication of secret-shared values. For more details, we refer to \cite{lindell2017framework}. In the case of binary fields, the MPC scheme \textbf{PsRepl2k} that we use was recently proposed by Eerikson et al.~\cite{eerikson2020use}; we evaluate the option with preprocessing for generation of the multiplication triples that is available in MP-SPDZ \cite{MP-SPDZ}. Note that in the three-party computation setting, the generation of the multiplication triples does not require any expensive public-key encryption operations.

\subsubsection{Quantization}\label{SEC:QUANT}
MPC based on secret sharing, as explained in Section \ref{SEC:ADDITIVESHARING} and \ref{SEC:REPLICATEDSHARING}, provides a mechanism to perform secure computations on \textit{integers} modulo $q$. The parameter values of a trained neural network, i.e.~the values in the filters in the convolutional layers, the weights on the dense layers etc., are natively \textit{real} numbers and need to be converted to integers. For this conversion process, we leverage existing research on quantization of neural networks. 
In deep learning, the conversion of floating-point (FP) data in the network to integers (INT) is studied as an effective way to shrink the model size 
and to accelerate computation, e.g.~on edge devices with limited memory and computational power \cite{yang2020training}. 
The use of quantization is growing in popularity in research on privacy-preserving deep learning as well, for instance in XONN \cite{riazi2019xonn}, where neural network parameters are restricted to take binary values $\{-1,1\}$, 
in Quotient \cite{agrawal2019quotient} with ternarized network weights in $\{-1,0,1\}$, and in SecureQ8 \cite{dalskov2019secure} where network weights are reduced to 8-bit integers. We adhere to the latter.

Quantization allows to represent a set of real numbers $\{\alpha_1,....,\alpha_n\}$ $\in$ $\mathbb{R}$ as a set of integers $\{a_1,....,a_n\} \in \mathbb{Z}_q$. In this work we use the 8-bit quantization method implemented in TensorFlow Lite,\footnote{\url{https://www.tensorflow.org/lite/performance/quantization_spec}} which was designed in the work of 
Jacob et al.~\cite{Jacob2018} and used previously in SecureQ8 \cite{dalskov2019secure}.
Let us define the dequantization function 
$$
\begin{array}{rrcl}
\mathsf{dequant}_{m,z}: 
& \{0,\ldots, 2^8-1\} & \rightarrow & \mathbb{R}   \\
& a & \mapsto & m \cdot (a-z)
\end{array}
$$
where $m \in \mathbb{R}$ is a scale and $z \in \{0,\ldots, 2^8-1\}$ is a zero-point. The quantization function $\mathsf{quant}_{m,z}: \mathcal{D} \rightarrow \{0,\ldots, 2^8-1\}$ 
with domain $\mathcal{D} =\{\alpha \in \mathbb{R}: -m \cdot z \leq \alpha \leq m \cdot (2^8-1-z)\}$ is then defined for an input $\alpha \in \mathcal{D}$ by picking the number $\alpha'$ in the image of $\mathsf{dequant}_{m,z}$ that is the closest\footnote{Breaking distance ties in favor of the smallest number.} to $\alpha$ and setting 
\begin{center}
$\mathsf{quant}_{m,z}(\alpha)= a$ such that $\mathsf{dequant}_{m,z}(a) = \alpha'$.
\end{center}

The quantization hyperparameters $m$ and $z$ are not the same across the entire neural network. The range of real values in the neural network may differ from one layer to the next. To ensure that all relevant real values are in $\mathcal{D}$, a pair $m$, $z$ is chosen ``per tensor'' in the neural network (in our case of 1-dimensional CNNs, this means ``per matrix'', see Section \ref{SEC:CNNARCH}). Suitable values for $m$ and $z$ are determined automatically with a post-training integer quantization algorithm on the trained CNN and artificially generated input data. 

Dot product is an important operation in CNNs, both for the convolutional layers and the dense layers. We use the same method as SecureQ8 \cite{dalskov2019secure} to compute dot products by using only integer arithmetic to sum the products of the vector elements (in $\mathbb{Z}_q$ for $q \gg 2^8$) and a single fixed-point multiplication to adjust to the proper scale for the output. Adding bias is handled by setting the scale of the bias representation to be the same as the scale of the output, and its zero-point to 0. Layers that only involve comparisons, such as RELU, can be directly implemented on the quantized values if they share the same scale and zero-point. 

When a fixed-point multiplication is performed, it is necessary to truncate the result by a number of bits equal to the number of bits that is used to represent the fractional part, so that the output does not use twice as many bits to represent the fractional part as the inputs. In the case of prime fields this is done using either the deterministic truncation protocol of Catrina and De Hoogh \cite{catrina2010improved} 
or the probabilistic truncation protocol of Catrina and Saxena \cite{catrina2010secure}. The protocol of Catrina and De Hoogh computes the exact truncation result, but needs to invoke a secure bitwise comparison protocol (which increases the overall runtime of the truncation) in order to verify if a modular reduction modulo $q$ occurred in a certain step of the secret-shared based protocol or not. We refer interested readers to the original paper for details: the deterministic truncation protocol is described as Protocol 3.3, and its building blocks are explained in Sections 2, 3 and 4 of that paper. In the case of the probabilistic truncation protocol of Catrina and Saxena, the probabilities that a number is rounded up or down are proportional to its distance to those bounds. 
The probabilistic truncation protocol eliminates the invocation of the underlying secure bitwise comparison protocol, and therefore  improves the efficiency. We refer interested readers to the original paper of Catrina and Saxena: the probabilistic truncation protocol is described as Protocol 3.1, and its building blocks are described in Section 2.

The probabilistic truncation affects negatively the accuracy of the secure classification as we will show in Section \ref{SEC:RESULTS}. In the case of binary fields, the truncation is done using the adaptations of the above deterministic and probabilistic truncation protocols that were introduced by Dalskov et al.~\cite{dalskov2019secure}. In the procedures in which the amount of bits to be truncated needs to be kept secret, we use the protocol of Dalskov et al.~\cite{dalskov2019secure} to perform deterministic truncation by a secret value.

We refer interested readers to \cite{dalskov2019secure,Jacob2018} for further details.

\subsubsection{Using an MPC scheme to securely classify} \label{SEC:SECCLASSIFICATION}
Classification of Alice's speech signal vector $X$ with Bob's model can, at a high level of abstraction, be thought of as the evaluation of a function $f(X,\theta)$ that depends both on Alice's input $X$ and on proprietary model parameters $\theta$ that were learned during training and that are only known to Bob. In the following description, we focus on the case of two-party computation for concreteness, but the case of outsourced three-party computation can be handled similarly. Designing a secure solution based on MPC for the classification comes down to representing the function that needs to be privately computed using the basic operations that are provided by the underlying MPC scheme (i.e., the addition and multiplication gates). Once this representation is found, the parties evaluate it gate by gate using existing procedures for private addition and private multiplication as explained in Section \ref{SEC:ADDITIVESHARING}. 
This classification is performed during the \textit{online phase} of protocol, consuming the necessary values that were generated during the \textit{offline phase}, i.e.~the multiplication triples that are needed for multiplication of secret-shared values, as well as the random values that are needed for Alice to secret share her speech signal vector $X$, and for Bob to secret share his model parameters $\theta$. During the secure classification process, Alice and Bob jointly go through the following steps:

\begin{enumerate}[leftmargin=*]
\item \label{STEP:INPUT} \textit{Input.} Alice secret shares her MFCC  vector $X$, and Bob secret shares his model parameters $\theta$ using the technique for secret sharing described in Section \ref{SEC:ADDITIVESHARING}. Since secret sharing is done over integers, this requires that Alice and Bob first convert their real valued numbers into integers, using either a fixed-point representation \cite{catrina2010secure} or, as we do in this paper, using a quantization scheme (Section \ref{SEC:QUANT}).

\item \label{STEP:CONV} \textit{Convolutional layer.} In this step, Alice and Bob need to compute a secret sharing $[\![Z]\!]$ of the output of the first convolutional layer, starting from the secret-shared input $[\![X]\!]$ and the secret-shared model parameters $[\![\theta]\!]$. As indicated in the pseudocode in (\ref{EQ:CODEFORLOOPS}) in Section \ref{SEC:CNNARCH}, to this end they need to perform Frobenius inner products (a generalization of dot product to matrices) and add bias terms. This boils down to performing multiplications and additions of values that are secret shared among Alice and Bob, namely Alice's speech signal vector $[\![X]\!]$ and Bob's model parameters $[\![W_i]\!]$ and $[\![b_i]\!]$ (which are part of $[\![\theta]\!]$). We refer to Section \ref{SEC:ADDITIVESHARING} and \ref{SEC:QUANT} for a description of how these operations are performed over secret shares.

\item \label{STEP:RELU} \textit{RELU activation layer.} In this step, Alice and Bob replace all negative values in $[\![Z]\!]$ by zeros. This is done directly in the quantized values using a secure comparison protocol derived from Catrina and De Hoogh \cite{catrina2010improved}, followed by a secure multiplication to either keep the original value or replace it by zero in an oblivious way. The secure comparison protocol of Catrina and De Hoogh  is based on protocols that securely compute: (1)
if a secret-shared value is equal to zero; and (2) if a secret-shared value is less than zero (which can be concluded from the most significant bit of the secret-shared value). For more information, we refer to the descriptions in Protocol 3.6 and 3.7 of Catrina and De Hoogh's paper, as well as the details of their underlying building blocks that are presented in Section 2, 3 and 4 of that paper. The comparison between two numbers $a$ and $b$ is then straightforwardly made using the above protocols and the difference $(a-b)$ or its negation. We refer to Table 2 of the original paper for details.

\item \label{STEP:POOLING} \textit{Average pooling layer.} Average pooling with a window size of P means that in every row in $Z$, each (non-overlapping) block of P adjacent elements is replaced by one cell, with the average value of the original block. The resulting matrix $Z'$ is smaller than the original matrix $Z$. The values in $Z$ are secret shared between Alice and Bob. To do average pooling, Alice and Bob first add the values in a block of $Z$ by adding their own shares of these values. Next Alice and Bob need to divide the resulting sum $[\![s]\!]$ by P, to yield the average. The window size P is a hyperparameter of the model that is known by Bob. Bob secret shares the value of hyperparameter P with Alice, similarly to how he shares the regular parameter values in step \ref{STEP:INPUT}. For secure division of $[\![s]\!]$ by $[\![\mbox{P}]\!]$, Alice and Bob use an iterative algorithm that is well known in the MPC literature \cite{catrina2010secure}. This is the protocol for secure division used for the experimental results in Section \ref{SEC:RESULTS}. There is room for optimization in the runtime if Bob is willing to leak the window size P to Alice. P is part of the neural network architecture, just like the size L of the filters in the convolutional layers. If both Alice and Bob know the value of hyperparameter P, then there is no need for them to execute a protocol for secure division, as they can simply multiply $[\![s]\!]$ by the constant 1/P without the need to communicate with each other.

\item \textit{More convolutional blocks.} Alice and Bob 
repeat steps \ref{STEP:CONV}-\ref{STEP:RELU}-\ref{STEP:POOLING} as many times as needed, depending on the neural network architecture.

\item \textit{Dense layer.} In a CNN, the output of the last convolutional block is flattened into a vector $\textbf{x}$ of
length $d$. Alice and Bob can each flatten their own shares of the values to construct $[\![\textbf{x}]\!]$. Next, $[\![\textbf{x}]\!]$ needs to be multiplied with a $d \times o$ matrix $[\![W_d]\!]$ that contains the weights of the dense layer, and a bias term $[\![b]\!]$ needs to be added.  $[\![W_d]\!]$ and $[\![b]\!]$ have already been provided by Bob as inputs in Step \ref{STEP:INPUT}. The output of the dense layer is a vector $\textbf{y}$ of length $o$. Alice and Bob jointly compute $[\![\textbf{y}]\!]$ by performing dot products and adding the bias term as explained in Section \ref{SEC:ADDITIVESHARING}-\ref{SEC:QUANT}.

\item \textit{Output layer.} The class label inferred by the CNN is the index corresponding to the largest value in $\textbf{y}$. In the final step, Alice and Bob obtain a secret sharing $[\![c]\!]$ of the class label by running a secure argmax protocol, which can be straightforwardly constructed using the above mentioned secure comparison protocol \cite{ToftThesis}.

\end{enumerate}



\section{Results}
\label{SEC:RESULTS}

Experimental setup: all benchmark and accuracy tests were completed on co-located F32s V2 (VM1) and F72s V2 (VM2) Azure virtual machines. We benchmarked our tests on two separate performance level machines to have a comparison of realistic runtimes today and into the future. A F32s V2 virtual machine contains 32 cores, 64 GiB of memory, and up to a 14 Gbps network bandwidth between each virtual machine.  The F72s V2 virtual machine represents computing power that could potentially be used more widespread in the future; it contains 96 cores, 144 GiB of memory, and a 30 Gbps connection speed between virtual machines. 


\subsection{Data preprocessing and model training}
\label{SEC:EVALMODELTRAIN}
We evaluated the proposed approach in a use case for emotion recognition from audio, using the 1,440 standard audio files from the RAVDESS data set \cite{livingstone2018ryerson}. 
Each audio file has a length of $\sim$3.5 sec, and is annotated with one of eight emotion labels: \textit{neutral} (57), \textit{calm} (113), \textit{sad} (134), \textit{happy} (131), \textit{fearful} (121), \textit{disgust} (60), \textit{angry} (130), and \textit{surprise} (64).
We extracted vectors of $m=40$ MFCC features from each audio file with the librosa library \cite{mcfee2015librosa}, with the default settings for all other parameters, and averaged them to obtain one 40-dimensional feature vector for each audio file. 

\begin{figure}
\begin{minipage}[c]{0.45\textwidth}
\begin{center}
\includegraphics[height=1.3in]{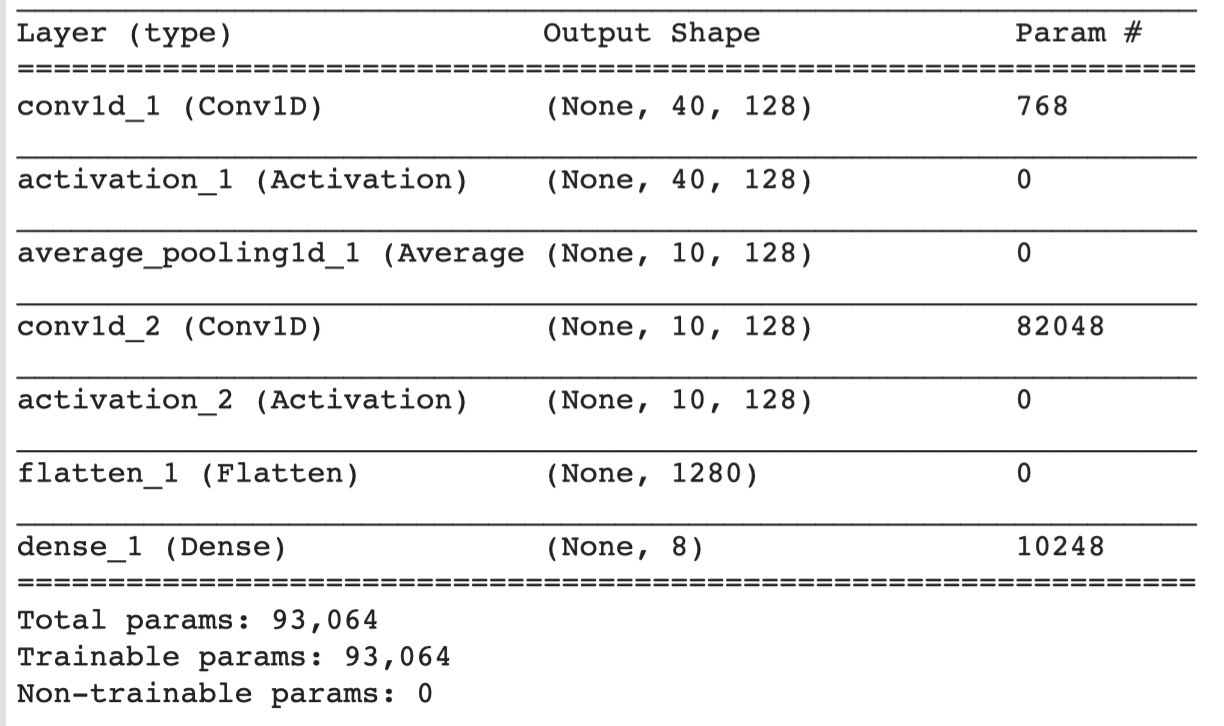}
\end{center}
\scriptsize{
Line (4), (11), and (12) in the code are only relevant for training, not for inference
}
\end{minipage}
\begin{minipage}[c]{0.55\textwidth}
\scriptsize{
\begin{tabular}{rl}
(1) & model = Sequential()\\
(2) & model.add(Conv1D(128,5,padding='same',\\
& input\_shape=(40,1)))\\
(3) & model.add(Activation('relu'))\\
(4) & model.add(Dropout(0.1))\\
(5) & model.add(AveragePooling1D(pool\_size=(4)))\\
(6) & model.add(Conv1D(128, 5,padding='same',))\\
(7) & model.add(Activation('relu'))\\
(8) & model.add(Dropout(0.1))\\
(9) & model.add(Flatten())\\
(10) & model.add(Dense(8))\\
(11) & model.add(Activation('softmax'))\\
(12) & opt = keras.optimizers.rmsprop(lr=0.00005,\\
& rho=0.9, epsilon=None, decay=0.0)
\end{tabular}
}
\end{minipage}
\caption{CNN architecture and Keras code snippet used to train the model. \label{FIG:ARCHITECTURE}}
\end{figure}

We used a CNN architecture with two convolutional blocks. Both convolutional blocks have RELU activation, and the first one has an average pooling layer for downsampling. The convolutional blocks are followed by a dense layer with softmax activation. A Keras\footnote{\url{https://keras.io/}} code snippet with more details is included in Figure \ref{FIG:ARCHITECTURE}. 
We hold out 33\% of the data as test data and train on the rest. The unquantized trained model achieves 72.11\% accuracy on the test data, while the quantized model has an accuracy of
70.32\%. These accuracies are state-of-the-art for speech emotion recognition on the RAVDESS dataset. Zamil et al.~\cite{zamil2019emotion} for instance report 70\% with a logistic model tree approach based on MFCC features; while Issa et al.~\cite{issa2020speech} report 71.61\% with a CNN approach based on MFCCs and Mel-scaled spectrograms.
We used TensorFlow Lite's post-training integer quantization\footnote{\url{https://www.tensorflow.org/lite/performance/post_training_integer_quant}} to convert all CNN model parameters to 8-bit integers.\footnote{Our source code can be accessed at} \url{https://github.com/KyleBittner/Private-Speech-Characterization}


\subsection{Secure inference}

To evaluate the accuracy and efficiency of the MPC schemes from Table \ref{TAB:MPCSCHEMES} for privacy-preserving emotion detection, we ran experiments with the quantized trained model from Section \ref{SEC:EVALMODELTRAIN} on the held-out test set. 
For the binary field $\mathbb{Z}_{2^k}$, a value $k=72$ was used, while for the prime field $\mathbb{Z}_p$ a prime number $p$ with bit length 64 was used.

Table \ref{TAB:RESULTS} contains accuracy and runtime results obtained on the two different configurations of the VMs in the case of two-party computation, while Table \ref{TAB:RESULTS2} contains similar data for the case of three-party computation. As mentioned above, the accuracy results were obtained by holding 33\% of the data out as test data. The classification runtimes are computed as an average over 10 inferences, and they include the time needed for both the offline and the online phases. As expected, the accuracy results are consistent across the VMs and the 2PC/3PC settings, while the runtime differs. 

First we observe that the accuracy results obtained with the deterministic truncation protocol are the same as the accuracy results in-the-clear (70.32\%, see Section \ref{SEC:EVALMODELTRAIN}), while the probabilistic truncation protocol causes a significant drop in accuracy to 64.38\%. These numbers are interesting by themselves: while Dalskov et al.~\cite{dalskov2019secure} write that the use of a probabilistic truncation protocol may hurt classification accuracy, to the best of our knowledge, we are the first to evaluate and measure this drop in accuracy experimentally on a real-life data set.

\begin{table}
$$
\footnotesize{
\begin{tabular}{|c|l|r|r|r|r|r|}
\cline{4-7}
\multicolumn{3}{c|}{} & \multicolumn{2}{c|}{\textbf{Active Security}} & \multicolumn{2}{c|}{\textbf{Passive Security}}\\
\hline
\textbf{VM} & \textbf{Truncation} & \textbf{Accuracy} & SPDZ2K & MASCOT & SEMI2K & SEMI  \\
\hline
\multirow{2}{*}{2 VM1s}
& Probabilistic & 64.38\% & 250.9 sec & 274.6 sec & 27.6 sec & 92.5 sec\\
& Deterministic & 70.32\% & 370.0 sec & 316.4 sec & 40.5 sec & 112.3 sec\\
\hline
\multirow{2}{*}{2 VM2s}
& Probabilistic & 64.38\% & 26.01 sec & 28.36 sec & 2.77 sec & 9.56 sec\\
& Deterministic & 70.32\% & 33.30 sec & 32.28 sec & 4.17 sec & 11.55 sec\\
\hline
\end{tabular}
}
$$
\caption{Accuracy and runtime results for privacy-preserving emotion detection in the dishonest-majority, 2PC setting in which Alice and Bob perform the privacy-preserving classification themselves. In each experiment, Alice and Bob were each run on a separate computing server (VM).} The accuracy results were obtained by holding 33\% of the data out as test data. The classification runtimes are computed as an average over 10 inferences.\label{TAB:RESULTS}
\end{table}

The absolute runtimes that we obtain are, even on the more modest VM, an order of magnitude smaller (better) than the runtimes reported for image classification in \cite{dalskov2019secure}. This is because our overall neural network architecture is far more compact; the fact that we choose to use a 1-dimensional CNN instead of a 2-dimensional CNN contributes to this gain in speed.
Beyond that, our runtime results are in line with what is reported in \cite{dalskov2019secure}. For the 2PC setting (Table \ref{TAB:RESULTS}) we observe the following:
\begin{itemize}[leftmargin=*]
    \item The probabilistic truncation protocol allows faster secure inferences than the deterministic truncation protocol. The price paid for this gain in speed, is a loss in accuracy (in our data set, a loss of $\sim$6\%).
    \item Among the MPC schemes for passive security, SEMI is 2-4x slower than SEMI2K.
    Among the MPC schemes for active security, the difference in runtime between SPDZ2K and MASCOT is minor (one slightly better with the deterministic truncation, the other slightly better with the
    probabilistic truncation).
     \item SEMI2K (passive security) is around 7-10x faster than SPDZ2K/MASCOT (active security). The price paid for this gain in speed is a weaker security setting, in which it is assumed that the adversary tries to gain additional information, but nevertheless follows the protocol specifications.
\end{itemize}

\begin{table}
$$
\footnotesize{
\begin{tabular}{|c|l|r|r|r|r|r|}
\cline{4-7}
\multicolumn{3}{c|}{} & \multicolumn{2}{c|}{\textbf{Active Security}} & \multicolumn{2}{c|}{\textbf{Passive Security}}\\
\hline
\textbf{VM} & \textbf{Truncation} & \textbf{Acc} & PsRepl2k & PsReplPrime & Repl2k & ReplPrime  \\
\hline
\multirow{2}{*}{3 VM1s}
& Probabilistic &64.48\% &  10.16 sec &  9.97 sec &  1.24 sec &  4.18 sec\\
& Deterministic & 70.32\% & 12.72 sec  &  12.44 sec &  2.06 sec & 4.86 sec\\
\hline
\multirow{2}{*}{3 VM2s}
& Probabilistic & 64.48\% & 1.35 sec & 1.32 sec & 0.15 sec & 0.52 sec\\
& Deterministic & 70.32\% & 1.61 sec & 1.58 sec & 0.26 sec & 0.60 sec\\
\hline
\end{tabular}
}
$$
\caption{Accuracy and runtime results for privacy-preserving emotion detection in the 
honest-majority, 3PC setting in which Alice and Bob outsource the privacy-preserving classification to be performed by three servers. The accuracy results were obtained by holding 33\% of the data out as test data. The classification runtimes are computed as an average over 10 inferences.\label{TAB:RESULTS2}}
\end{table}

The protocols in the three-party outsourced computation setting with honest majority execute between 16x and 29x faster than their counterparts in the two-party computation setting. This is expected given the performance differences between state-of-art MPC protocols in the 2PC with dishonest-majority and 3PC with honest-majority settings. Beyond that, we have that

\begin{itemize}
    \item Among the MPC schemes for active security, PsReplPrime (which uses a prime field) performs slightly better than PsRepl2k (which uses a binary field) in all tests.
    \item On the other hand, among the MPC schemes for passive security, Repl2k outperforms ReplPrime in all tests, running around 2-3x faster.
    \item Repl2k executes around 6-9x faster than PsReplPrime.
\end{itemize}

Considering passive security in both the 2PC and 3PC settings, performing the secure classification using computations on a binary field is far more efficient than using a prime field. 
On the other hand, in the active-security setting, the secure classification achieves a comparable running time on both binary and prime fields, the winner depending on the number of parties running the MPC scheme and the type of truncation. Note that, in the passive-security setting the overall procedures required for performing a multiplication of secret-shared values are far less complicated than in the active-security setting, and in the active-security setting those procedures are more complicated in the case of binary fields.

Towards deployment in a real-time privacy-preserving speech classification application, the 3PC setting with three semi-honest computational servers is a very viable option (Table \ref{TAB:RESULTS2}). The gain in speed compared to the 2PC setting stems from the use of cryptographic protocols that leverage the availability of three instead of only two players to secret share the values with, and the removal of the need for expensive public key encryption, rather than the availability of more hardware in the form of a third server. It is important to stress that, since the three servers only receive shares of Alice's and Bob's information, \emph{the servers do not learn anything} about the speech signal nor the trained model parameters. This holds true as long as not more than one of the three servers is corrupted. The 3PC setting is a good fit for applications where the user (Alice) and the application developer (Bob) have access to three reliable computational servers in the cloud, and the application developer wants to offer a speech classification service without becoming liable for invading the user's privacy.

In settings where there is no configuration available of three computational servers with an honest majority, and where each party can only trust itself, one can resort to the MPC schemes from the 2PC setting at a higher runtime cost (see Table \ref{TAB:RESULTS}). These may be suitable for sensitive applications where real-time speech classification is not a requirement, such as healthcare applications or empathy based AI systems where one can afford several seconds of even half a minute to detect a disease or the user's general mood in a privacy-preserving manner.

\section{Conclusion}\label{SEC:CONCLUSION}
In this paper, we have presented the first privacy-preserving approach to deep learning based speech classification that is provably secure. To this end, we have proposed the first  application of privacy-preserving classification with 1-dimensional CNNs based on Secure Multiparty Computation (MPC).
In terms of privacy, MPC is very reliable: other than the result of the classification (which can be selectively revealed
to the model owner, data owner, or a third party depending on the application), no information about the speech signal or the trained model parameters is leaked to any participant of the protocol. When performing oblivious speech classification, the price paid for keeping the data and the model private, is an increase in computational cost and runtime. Our results answer a question that has remained open in the literature thus far, namely whether MPC based protocols are efficient enough to enable highly accurate real-time speech classification as would be needed for instance for digital voice assistants such as Apple's Siri, Amazon's Alexa, Google Home, and Microsoft's Cortana. Our results show that this is clearly within reach.

In our experiments for a passive-security setting, i.e.~with semi-honest parties who follow the instructions of the cryptographic protocols, an audio file of 3.5 sec is classified with high accuracy in 0.26 sec, and in 0.15 sec with lower accuracy. These results were obtained with a CNN that we optimized for high accuracy as well as high efficiency in the MPC setting, through deliberate design choices in the CNN architecture, and the use of quantization. We ran the protocols in MP-SPDZ, an existing framework for MPC that is not optimized in any specific way for speech classification. That means that, in addition to the optimization efforts we made in this paper on the machine learning side, there is room to bring the secure inference runtimes down even further by optimizations on the MPC side, for instance by replacing the division algorithm in the average pooling layer by multiplication with a constant.    

The fastest results mentioned above are obtained when Alice and Bob outsource the computations to three semi-honest servers (3PC). As long as these servers do not collude with each other, they do not learn anything about Alice's speech signal or about Bob's trained model parameters. We have also included scenarios with stronger security assumptions in our study, namely, in increasing order of runtime: malicious adversaries with an honest majority (3PC), semi-honest adversaries with a dishonest majority (2PC), and malicious adversaries with a dishonest majority (2PC). Actively secure protocols remain secure even if one of the parties is a malicious adversary who deviates from the protocol specification. This makes these protocols most suitable for sensitive applications, even if they come at a notably higher computational cost.


\section*{Acknowledgements}
The authors would like to thank Marcel Keller for making the MP-SPDZ framework available, and for his assistance in the use of the framework.

The authors would like to thank Microsoft for the generous donation of cloud computing credits through the UW Azure Cloud Computing Credits for Research program. Beyond that, this research did not receive any specific grant from funding agencies in the public, commercial, or not-for-profit sectors.

\bibliography{references,crypto,abbrev0}

\end{document}